# Photon-recycling dielectric laser accelerator

Changying Li[1], Li Zhang[1], Dingguo Zheng[3], Xiaoping Liu[1], Yiming Pan[1,2]

1. School of Physical Science and Technology, ShanghaiTech University, Shanghai 200031, China
2. Center for Transformative Science, ShanghaiTech University, Shanghai 200031, China
3. Department of Electrical Engineering Physical Electronics, Tel Aviv University, Ramat Aviv 6997801, Israel

## Abstract

We propose a photon-recycling dielectric laser accelerator (DLA) system based on silicon photonic device. Our DLA system employs guided electromagnetic waves as a primary energy source, modulated to inject into the electron-light interaction region to accelerate or modulate electron beams and recycled the energy for the next round-trip. Long-distance acceleration takes place as electrons interact with the pre-modulated light field. Our loop recycles post-interaction light field, enabling photons reuse across successive cycles. To optimize the interaction process, we developed an adaptive algorithm to refine waveguide structures, and identified an "optimal waveguide accelerator" with superior performance on our dataset. We find that the optimized DLA loop only requires low-power light injection to sufficiently sustain high acceleration gradients for continuous electron beams. Under optimal electron beam intensity, the system achieves exceptionally high photon utilization, ensuring that nearly all injected light power transferred to electrons. Using spectral analysis, we demonstrate that the optimal waveguide also operates as an electron energy filter, selecting and manipulating phase-matched electrons over a broad energy range, even for quantum electron wavefunction shaping. Our photon-recycling DLA setup is not only suitable for low-energy beam accelerators, but also offers versatility as a beam filter or a narrow energy selection combined with other optical elements, the total setup can be further applied to explore free electron quantum optics engaging with the advancing field of photonic integrated circuits.

The interaction between light and electrons is ubiquitous, with decades of advancements giving rise to numerous technologies and achievements. Laser-driven dielectric laser accelerators (DLAs) are one of excellent tools for exploring light-electron interactions at the nanoscale. Sub-wavelength structures induce near-field effects, which enable accelerating gradients that are 1 to 2 orders of magnitude higher than those of conventional radio-frequency (RF) accelerators for both relativistic and sub-relativistic electrons. Recently, inverse Cherenkov radiation accelerators utilizing quartz, known for its high breakdown voltage, have demonstrated accelerating gradients exceeding 1.6 GeV/m, with characteristic feature sizes ranging between 100 and 250 microns [1, 2]. In contrast, free-space inverse Smith-Purcell DLAs have minimized system dimensions to the nanoscale. These DLAs achieve accelerating gradients from 200 MeV/m to 1 GeV/m for sub-relativistic electrons [3-11], and from 300 MeV/m to over 850 MeV/m for relativistic electrons [12-16], with interaction lengths spanning tens [3, 4, 15, 17] to hundreds of microns [10, 16]. Notably, dual-pillar grating geometries are versatile, supporting acceleration, deflection, or focusing modes from sub-relativistic to fully relativistic speeds, making them widely used in demonstrations and applications. However, the optimal bunch charge per pulse for these DLAs is generally in the femtocoulomb (fC) range [1, 2, 7, 13], with only a small fraction of photons participating in electron acceleration. Most photons are scattered into space, while some are absorbed by materials (typically silicon), generating heat that causes non-uniform temperature distributions. These thermal-optical effects, in turn, influence near-field electron-light couplings.

Enhancing photon utilization efficiency in DLAs involves many aspects, such as increasing the injected current, extending the interaction length, which holds the potential to revolutionize the state-of-the-art acceleration schemes. For instance, multichannel DLAs [13] can boost the total electron throughput by one order of magnitude, while cascading the structures [18-23] can further extend the interaction length by more than an order of magnitude. However, the cascading process is limited by electron beam divergence and coherent light source cascade. Alternating-phase focusing Inverse Smith-Purcell DLAs [10, 22, 23] effectively address this by suppressing divergence, thus enabling cascading. Regarding light source constraints, solutions include delaying identical and phase-locked laser pulses or split-beam pulses [1, 18-21], and using gratings to generate front-tilted laser pulses [6, 10, 14, 22] have been proposed. These methods, however, require high precision in laser delivery systems and significantly increase system size, hindering the integration of DLA systems. Additionally, in Inverse Smith-Purcell DLAs, non-zero angle laser incidence introduces phase-dependent deflection forces on electrons [14, 23], whereas co-propagation of the light and electrons minimizes transverse deflection forces. By employing traveling-wave structures—such as flat interfaces[24], gratings and photonic crystals [25-27], ring resonators [28-30], dielectric-lined waveguides [31-37], and slot waveguide [38, 39] —the coupling length can be maximized, thereby enhancing the feasibility of cascading design of DLAs [1, 36, 37].

To practically minimize an accelerator on a chip, it is essential to integrate the accelerator components with waveguide structures. Silicon, with its excellent thermal conductivity, electrical properties, and compatibility with the advanced microfabrication techniques, has been extensively employed in integrated photonics. Various low-insertion-loss beam splitters [40], polarization devices [41, 42], and high-efficiency fiber-to-chip couplers[43, 44] have been developed, making it feasible to integrate optical components with particle accelerator systems. In the Inverse Smith-Purcell DLA, guided waves can replace free-space lasers as a stable source [45, 46]. High-performance beam splitters and low-loss waveguides enable the design of a laser delivery system with uniform splitting and precise delay, suitable for cascading systems [47]. Moreover, the development of microfabrication makes strongly interacting travelling-wave accelerators possible. Compared to dielectric-lined waveguides [31-36], nano-waveguide DLAs [38, 39] can significantly compress guiding modes, while microscale micro-ring DLAs [28-30] feature high-Q resonance structures. Both approaches effectively increase the field enhancement factor (the ratio of maximum electric field in the DLA to the input field), maximizing coupling strength. Recent studies demonstrate that waveguide DLAs can achieve acceleration gradients as high as 1.68 GeV/m with interaction lengths of no less than 50 $\mu$m [39]. Similarly, micro-ring DLAs have reached gradients of 1.63 GeV/m, with post-cascade energy exchange exceeding 100 keV [29]. However, the optimal bunch charge for waveguide DLAs remains at sub-fC levels, and the optical-to-beam conversion efficiency is below 1% [38]. In single-pass DLAs, most photons are scattered or absorbed by waveguide structures, converting into heat, which causes significant refractive index fluctuations and disrupts phase matching conditions.

In this work, we designed an on-chip racecourse DLA setup (see Fig. 1a), inspired by synchrotrons storage rings. This DLA setup retains many advantages of waveguide-based DLAs while incorporating a recycling process to maximize photon utilization. The accelerator employs a nonlinear tapered nano-waveguide, ensuring phase matching between electron beams and photons in a strong electric field. At the end of the acceleration area, a curved waveguide collects the remaining photons and recycle them by directing them into the combiner for the next acceleration cycle. A negative-feedback combiner merges the recycled photons with the light injected from the grating coupler and fiber. The combined beam is then output and mode-corrected by a linear tapered waveguide before reinjection into the accelerator. This forms a ring resonator, where the input light is amplified through resonance, significantly reducing the demand for high-power lasers and high-damage-threshold gratings. This system is realized on a silicon-on-insulator (SOI) platform and operates at the 1550 nm telecommunication wavelength. Numerical simulations show that the photon-recycling DLA requires only 17.2 mW of continuous input optical power to sustain 100 mW of operating power in the loop, with a relaxation time of fewer than 20 cycles (approximately 10 ns), indicating a low-cost but high-efficient acceleration process. The accelerator continuously accelerates electrons from 22.9 keV to 198.4 keV over an interaction length of 13.9 mm, achieving an average acceleration gradient of 12.6 MeV/m and a peak gradient of 20 MeV/m, with a field enhancement factor of 4.8. About 70% of the energy is recycled into the next cycle. Moreover, at a beam current of 0.285 µA, our photon-recycling DLA demonstrates a photon utilization efficiency nearing 99.8%, and reduce thermal effects significantly via the recycling process, as compared to conventional single-pass DLAs.

**Non-linear tapered accelerating waveguide**. The proposed accelerating waveguide consists of a high-refractive-index material fabricated on a low-refractive-index substrate. When optically excited in the transverse magnetic (TM) mode, the waveguide exhibits an exponentially decaying electric field component above its upper surface. This leakage field enables the interaction between electrons and light, thereby affecting the electrons' motion. To achieve phase matching, we taper the waveguide along the propagation direction. Consider a continuously accelerated electron, referred to as the target particle, its motion is governed by the relativistic form of Lorentz equation. The phase-matching condition is therefore expressed in the following integral form (see Methods):

$$\left(\frac{1}{n_{\text{eff}}}\right)^2 = \beta^2 = 1 - \left[\frac{\left(E_{kin,0} + e \int_0^x E_x(x')dx'\right) + E_0}{E_0}\right]^{-2}. \tag{1}$$

Both leaked field and effective refractive index depend on the waveguide profile:

$$E_x = E_x\big(w(x), h(x)\big) = E_x(x), \qquad n_{\text{eff}} = n_{\text{eff}}\big(w(x), h(x)\big) = n_{\text{eff}}(x).$$

To design a phase-matched waveguide, we must determine the appropriate width $w(x)$ and height $h(x)$ that satisfy Eq. (1), which is a tricky task. Alternatively, we address the inverse design problem by finding the reciprocal relation between $x$, and ($h$, $w$), that is, for a given $x$, what should ($h$, $w$) be? Or conversely, for a given ($h$, $w$), what should $x$ be?

The design procedure is outlined in the top subgraph of Fig. 2. To solve the inverse problem, we discretize the continuously varying waveguide footprint $\big(w(x), h(x)\big)$ into a series of points $(w_i, h_i)$. A Finite Difference Eigenmode (FDE) solver is used to compute the effective refractive index $n_{\text{eff},i}$ and the leaked field intensity $E_{x,i}^{(\text{FDE})}$. An energy normalization approach is then applied $E_{x,i} = E_{x,i}^{(\text{FDE})} \sqrt{P_i/P_i^{(\text{FDE})}}$, ensuring energy conservation. The refractive index $n_{\text{eff},i}$ is translated into electron kinetic energy $E_{kin,i}$ using the phase-matching equation (Eq. M1). Once the discrete dataset is processed, a stepwise iterative algorithm, depicted in the bottom subgraph of Fig. 2, is applied. Assuming the target electron interacts with the optical field at the $i$-th cross-section, the algorithm calculates the distance $\Delta x_i$ to the next coupling cross-section $i$+1 based on the electron's dynamics and the waveguide field distribution. By summing these distances cumulatively, each position of the discretized cross-section $x_i$ is determined, thereby defining both the waveguide profile and the electron's motion during the interaction.

To thoroughly analyze the characteristics and performance of waveguide accelerators, we use silicon-based waveguides as a case study to build datasets and examine general trends. Current DLAs typically rely on ultrashort pulse as an energy source, achieving peak powers exceeding hundreds of watts to generate acceleration gradients in the order of GeV/m. However, this approach introduces challenges, such as waveguide damage, dispersion broadening, and nonlinear effects. In contrast, our proposed waveguide DLA design leverages optical loops for photon recycling and explores the feasibility of excitation using low-intensity continuous light (100 mW). By neglecting Coulomb repulsion between electrons, we focus on scenarios involving only a small number of electrons or a single electron. Thus, we assume $P_{loss} = 0$, and

$$P_{i-1} = P_i = P_0, \qquad E_{x,i} = E_{x,i}^{(FDE)} \sqrt{\frac{P_0}{P_i^{(FDE)}}}.$$

To evaluate the performance of waveguide accelerators, thousands of discrete data points $(h_i, w_i)$ are generated within the range of $50\,\text{nm} \leq h \leq 5\,\mu\text{m}$ and $50\,\text{nm} \leq w \leq 10\,\mu\text{m}$, and we calculated the corresponding effective refractive indices and leakage electric field strengths $(n_{\text{eff},i}, E_{x,i})$ for each point using the *FDE* solver and the energy normalization formula Eq. (M4). The effective refractive index describes the propagation properties of the guiding mode, while the leakage electric field is tied to the electron acceleration. Using the phase-matching equation Eq. (M1), we converted the waveguide's effective refractive index $n_{\text{eff},i}$ into the electron's kinetic energy $E_{kin,i}$, forming the dataset $(E_{kin,i}, E_{x,i})$, as shown in Fig. 3a. Each data point $(E_{kin,i}, E_{x,i})$ corresponds to a unique cross-section $(h_i, w_i)$, implying that an electron with kinetic energy $E_{kin,i}$ can be phase-matched and accelerated by an SOI silicon waveguide of $(h_i, w_i)$, which provides a leakage field with strength $E_{x,i}$ for acceleration. Furthermore, Fig. 3a illustrates several data points where the kinetic energy $E_{kin,i}$ is similar, but the field strengths $E_{x,i}$ differ substantially. These points reveals that for electrons at a given kinetic energy $E_{kin,i}$, multiple waveguide sizes $(h_i, w_i)$ can achieve phase matching and particle acceleration. However, different sizes generate different electric field strengths, resulting in considerable variations in acceleration gradients. Consequently, among the waveguides that fulfill phase matching, there exists one (or a set) that yields optimal acceleration, evident from the maximum leakage electric field strength $E_x$ and the highest acceleration gradient at a given kinetic energy $E_{kin}$.

By connecting different data points in Fig.3a, we construct "trumpet waveguides" with fixed thickness *h* and "slide waveguides" with fixed width *w*. The acceleration characteristics of these two types of waveguide accelerators will be analyzed in detail in the supplementary materials. Further simulations reveals that all the data points in Fig. 3a lie within a specific range, bounded by an envelope curve (the black dashed line of Fig. 3a) such that all points are below this curve. On this envelope curve, the electric field strength $E_x$ is maximized for a given kinetic energy $E_{kin}$. Thus, the "optimized waveguide" corresponding to the envelope curve should possess the highest acceleration gradient. In other words, no other waveguide can surpass the acceleration gradient of the optimized waveguide. Although calculating the optimized waveguide profile is challenging, approximating it is achievable. We extracted all data points at the upper edge of Fig. 3a $(E_{kin}, E_x)_{opt}$ and mapped them to their corresponding cross-sections $(h, w)_{opt}$. Due to minor data fluctuations, we performed a fitting procedure on $(h, w)_{opt}$ (Fig. 3c), which yields curves for the optimized height (blue area) and width (red area) as a function of $x$,

$$h - h_0 = \alpha_h e^{-\gamma_h(x-x_0)}$$
$$w - w_0 = \alpha_w (x - x_0)^{-\gamma_w},$$

and the relationship of the optimized waveguide profile (the subgraph of Fig. 3c),

$$\ln\left(\frac{h - h_0}{h_\infty - h_0}\right) = -\alpha(w - w_0)^{-\gamma}, \qquad \gamma > 0.$$

Both $h(x)$ and $w(x)$ decrease from larger initial values at a faster rate, then slow down as they approach smaller dimensions, with the width shrinking more rapidly. Fig. 3d shows how the optimized electric field changes during acceleration, along with the accelerated electron's kinetic energy curve. From the curve, we see that the optimized waveguide accelerates electrons from 22.9 keV to 198.4 keV over 13.9 mm, yielding an average acceleration gradient of 12.6 MeV/m and a peak gradient of 20 MeV/m. Initially, because the large waveguide strongly confines the guiding mode, the optimized electric field strength is low and changes slowly. As the waveguide cross-section decrease, confinement weakens and the leakage field gradually rises, peaking at 9 mm (with $w$=250 nm and $h$=299 nm). Further reduction in cross-section transfers the light intensity from the silicon waveguide to the $SiO_2$ substrate, lowering the leakage field strength. Theoretically, the optimized waveguide can reach an acceleration up to 198.4 keV, but this comes at the cost of having the optical field mainly in the $SiO_2$ box, complicating subsequent photon recycling. Therefore, for practical use, we strongly recommend truncating the tail section to below 195 keV and increasing the initial electron energy to improve overall acceleration. To this end, the optimized waveguide provides a long light-electron interaction distance, effectively accelerating incident electrons from below 30 keV to nearly 200 keV.

Notably, in our inverse design, under the assumption $P_i = P_0$, all parameters except for $x_i$, are independent to the previous $i$-th terms. Thus, even if we truncate the optimized waveguide, for instance, an acceleration segment from 5.45 mm to 13.55 mm, corresponding to the acceleration of electrons from 50 keV to 195 keV, this truncated portion remains optimal, providing the highest acceleration gradient for electrons in the same kinetic energy range, indicating a feasibility of our DLA design for arbitrary electron beams.

**Photon-recycling system and high photon-utilization accelerating waveguide.** In this section, we incorporate the optimized waveguide described previously into a full racecourse shown in Fig. 1a to create a photon recycling DLA system. For simplicity, we assume that all waveguides are lossless, so that only the combiner and the accelerating waveguide affect energy variations. The combiner serves as the photon source in the loop, while the accelerating waveguide functions as the energy exchanger, responsible for photon loss and gain. Fig 4a shows energy changes of the light field at key nodes in the loop. Laser field from a fiber, enters one side of the combiner and coheres with the recovered field from the other end. After coherence, most of light exits from the combiner output, while a small amount scatters at the sides. We can also add a branch to extract a small fraction of light for diagnosis, providing adaptation and feedback on the injected light source. The output intensity of the combiner is described by its transmission coefficient, which depends on the ratio of two input beam intensities at the ports (the subgraph of Fig. 4b):

$$T \triangleq \frac{P_{out}}{(P_{inj} + P_{rec})} = T(r),$$

with $r = P_{rec}/P_{inj}$ and $T(r) = T(r^{-1})$. From the transmission characteristic curve, we see that when both input beams have similar intensity, the intensity summing effect in the combiner approaches 1. As the intensity difference grows, scattering loss in the combiner increases. Under single-port input, only 50% of the light energy emerges from the output port. If no external input is injected, 50% of the loop energy is lost in each cycle, and after 20 cycles, the light intensity in the loop diminishes to a negligible level (left graph of Fig. 4b). However, if energy is continuously injected, the loop intensity stabilizes at a constant value, regardless of its initial energy, within about 20 cycles (right graph of Fig. 4b).

In the steady-state loop, the power injected from the combiner directly correlated with the power loss in the accelerating waveguide, representing the transfer of energy from light to electron (i.e., laser acceleration). Hence, we define this energy transfer as the "accelerating loss", and express the change in waveguide power before and after acceleration as $P_{rec} = P_{out} - P_{loss}$. Then, from the transmission characteristic curve of the combiner, we have:

$$\begin{aligned}\frac{P_{loss}}{P_{out}} &= 1 - \frac{P_{rec}}{P_{out}} = 1 - \frac{r}{T(r) \cdot (r+1)}, \\ \frac{P_{inj}}{P_{out}} &= \frac{P_{inj}}{T \cdot (P_{inj} + P_{rec})} = \frac{1}{T(r) \cdot (r+1)},\end{aligned} \quad (2)$$

We found that the injected energy and the accelerating loss ultimately depends on the parameter $r$, producing the relation between the normalized $P_{loss}$ and normalized $P_{inj}$ (blue line in Fig. 4c). The curve indicates that, in the low-loss region, the normalized $P_{inj}$ is below 1, with a minimum of 0.172. In other words, when only a few electrons is accelerated, injecting around 17.2% of the required light intensity is sufficient to maintain the loop at a constant out power $P_{out}$, while the remaining 82.8% is recovered from the previous cycle. This loop greatly reduces the input power and energy demands, confirming the highly efficient photon-recycling capability of our DLA design.

Accelerating a single electron means that only a small fraction of the many injected photons will interact with the electron. To quantify the fraction, we define the photon efficiency ($\eta$) as $\eta \triangleq P_{loss}/P_{inj}$. A higher $\eta$ indicates that more photons transfer energy to the electrons. Using this definition, we plot the relationship between photon efficiency and accelerating loss (red line in Fig. 4c). The Loss-η curve shows that, when accelerating a single electron, photon utilization is initially low because only a few photons can interact with electron at a given time. However, as the electron beam current increases, the rate of energy transfer from photons to electrons grows, driving the photon utilization upward until it reaches a maximum. This maximum appears near 0.5 normalized $P_{loss}$, corresponding to a photon efficiency of 99.78%. Specifically, if the phase-matched waveguide in the loop can transfer 50.03% of photon energy to the electron beam at a given beam current $I$, only 50.14% of the photons need to be injected to sustain stable energy levels in the loop, with 99.78% of the injected photons interacting with the electron beam.

Notably, we aim to design an accelerating waveguide that achieves phase matching with 50% energy exchange. Initially, we consider only the one-way transfer of energy from photons to electrons. According to energy conservation, the incident current $I$ is related to the accelerating loss by:

$$P_{loss} = \frac{I}{e}\Delta E_{kin} \qquad (3)$$

where $\Delta E_{kin}$ is the gain of the electron energy. Using Eqs. (M5-M8), we recalculate the optimized waveguide shown in Fig. 3. We design a phase-matched accelerating waveguide with a 100 mW input and an energy exchange of 50.03 mW, corresponding to $I$=0.285 $\mu$A. The results, presented in Fig. 4d, indicate that as the high-efficiency waveguide converts photon energy to the electron beam, the light intensity drops correspondingly along the propagation distance. Compared to the high-gradient accelerating waveguide that neglects accelerating losses, the energy decay leads to a lower electric field, slightly reducing the overall acceleration gradient and lengthening the accelerator. In total, the length of the waveguide accelerator increased by 12.9%, the acceleration gradient decreased to 88.6% of its expected maximal value, yet the photon efficiency rises from near 0% to 99.78%.

**Spectral analysis of optimized waveguide accelerators.** In this section, we perform a computational analysis of the electron gain spectrum in the optimized waveguide. Notice that electrons are modeled as point particles and we assume that the waveguide accelerates only one electron at a time, effectively eliminating Coulomb interactions between them. Fig. 5a illustrates the electron gain with varying incident energies and phases, revealing three key characteristics. First, the peaks in Fig. 5a are distributed across a broad energy range, suggesting that electrons with incident kinetic energies between 22.9 keV and 198.4 keV can be efficiently accelerated by the waveguide, provided there is an appropriate delay time. Second, each gain peak exhibits a characteristic width of less than 5 meV, with most of the peak values concentrated around 198.4 keV, indicating a high spectral resolution. Additionally, it means that the energy exchanges outside of the gain peaks are negligible. Third, as the incident kinetic energy increases, the gain peaks shift progressively toward longer delay times, ultimately forming a curve that spans over 100 keV. Each point on this curve, denoted a $(E_{kin,0}, t)$, represents an electron with incident kinetic energy $E_{kin,0}$ and delay time $t$, which is efficiently accelerated by the waveguide and exits with higher energy. We note that the delay time is defined as the time difference between the incident electron and the peak of the light wave at the initial position of the waveguide $x = 0$, corresponding to the electron's incident phase $\varphi_0 = \omega t$.

The underlying causes of these phenomena are attributed to the extended interaction length. First, as the phase velocity of photons increases along the waveguide, higher-energy electrons attain a velocity-matching condition with the photons at some point along the propagation path. At this moment, if the phase difference between the electron and photon is zero, and the electron is at the peak of the wave, the partical will be accelerated at the expected rate and maintain phase matching thereafter. Ultimately, the electron exits the interaction region with the anticipated kinetic energy. Conversely, if a phase mismatch exists at the point of velocity matching, the condition is disrupted immediately. Phase mismatches can accumulate during the acceleration process, leading to a position-dependent phase difference between the electrons and waveguide modes, which ultimately causes the average acceleration gradient to drop to zero (upper-left panel of Fig. 5b). In this case, unmatched electrons exit the acceleration waveguide region with their initial kinetic energy. Precise phase matching is governed by the delay time, making accurate timing critical for effective injection into the acceleration system. If an electron with kinetic energy $E_{kin,1}$ cannot be effectively accelerated when injected at time $t = 0$, our analysis indicates that by adjusting the injection time, the phase difference $\varphi_1$ at the velocity-matching position can be modulated. Eventually, an optimal injection time $t = t_1 < T$ will allow $\varphi_1 = 0$, ensuring effective acceleration under phase-matching conditions. This analysis explains the shift of gain peaks towards longer delay times at higher energies and demonstrates that each energy corresponds to a specific injection time that facilitates acceleration.

Finally, we observe that the acceleration waveguide generates only two distinct states of output electrons for the majority of incident electrons, analogous to a "band-stop filter". For a waveguide with a specified incident kinetic energy $E_{kin,0}$ and an output kinetic energy $E_{kin,out}$, electrons with energies below $E_{kin,0}$ or above $E_{kin,out}$ experience no acceleration, as the waveguide is "transparent" to them. By adjusting the delay time and considering the possibility of cyclic acceleration, electrons with energies between $E_{kin,0}$ and $E_{kin,out}$ can be efficiently accelerated to the target output energy $E_{kin,out}$. Therefore, the output electrons exist in only two states: those with energies lower than $E_{kin,0}$ and those with energies greater than or equal to $E_{kin,out}$.

These two states can be easily distinguished due to their energy difference. Consequently, the acceleration waveguide can function as a spectral filter, selectively allowing electrons with energies below or above a certain threshold to pass. Furthermore, if the incident electron energy lies between $E_{kin,0}$ and $E_{kin,out}$, after cyclic acceleration, the output energy should be concentrated within a few eV of $E_{kin,out}$.

**Conclusion.** The photon-electron phase-matching equation defines the physical principle that the acceleration waveguide must adhere to; however, solving for the waveguide profile from this equation directly is complex. By adopting an inverse-design approach, we simplified the phase-matching equation into a first-order integral equation, which was discretized to yield a set of algebraic equations. We systematically analyzed the characteristics of the silicon waveguide accelerator and optimized it to achieve the maximum acceleration gradient. We examined the electron dynamics on the optimized waveguide, and found that by controlling the delay time of the electron, we can adjust its kinetic energy to fall within the effective range of the waveguide. Our waveguide accelerator offers a novel approach for sub-relativistic electron acceleration by providing an effective method to generate an accelerating field that can remain synchronously extended with the accelerated electrons.

In combination with the optimized waveguide, we designed a DLA system with energy recovery by utilizing a Y-branch as a combiner. The system is capable of reusing most of the residual light from the preceding acceleration cycle in the subsequent cycle. This requires only low-power light injection to maintain energy stability across the loop, substantially reducing the injection requirements. Our simulations show that appropriate injection of electron current can substantially increase photon utilization, thereby enabling highly efficient photon usage. With precise control over the delay time between the electron and guiding wave peaks, a waveguide with an extended interaction length can effectively accelerate electrons whose kinetic energy lies between the designed incident and output energies, ultimately causing them to exit the waveguide region with kinetic energy closely aligned with the designed output value.

Added that, to fully leverage the properties of the acceleration waveguide, we employed Si with a high refractive index as the core material. The inverse-design approach and photon-recycling design are also applicable to other materials used in integrated photonic platforms, such as silicon nitride (SiN) with a high damage threshold. The photon-recycling acceleration waveguides utilizing other materials should exhibit comparable characteristics. In this study, we provided a comprehensive analysis and discussion of the photon-recycling DLA, thereby acting as a complete design guideline.

**Acknowledgement:**

We thank the discussions with Bin Zhang. Y.P. acknowledges the support of the NSFC (No. 2023X0201-417-03). X.P thanks the fund of the ShanghaiTech University (Start-up funding) and Science and Technology Commission of Shanghai Municipality (21DZ1101500).

The authors declare no competing financial interests.

Correspondence and requests for materials should be addressed to Y.P. (yiming.pan@shanghaitech.edu.cn), X.P. (liuxp1@shanghaitech.edu.cn) and D.Z. (zhengdingguo@tauex.tau.ac.il)

**Method**

Here, we given the details of the calculation procedure. Before going any further, we simplify Eq. (1) to compute the electron's kinetic energy at the $i$-th cross-section:

$$E_0\left[1-\left(\frac{1}{n_{\text{eff},i}}\right)^2\right]^{-\frac{1}{2}} - E_0 = E_{kin,0} + e\int_0^{x_i} E_x(x')dx' = E_{kin,i}. \quad (M1)$$

Thus, the kinetic energy difference between adjacent cross-sections is expressed as:

$$E_0\left[1-\left(\frac{1}{n_{\text{eff},i}}\right)^2\right]^{-\frac{1}{2}} - E_0\left[1-\left(\frac{1}{n_{\text{eff},i-1}}\right)^2\right]^{-\frac{1}{2}} = E_{kin,i} - E_{kin,i-1} = e\int_{x_{i-1}}^{x_i} E_x(x')dx'. \quad (M2)$$

Eq.(M2) shows, to determine $x_i$ of each section, we require the effective refractive index ($n_{\text{eff}}$) and field intensity ($E_x$).

Employing a *FDE* solver, the effective refractive index $n_{\text{eff},i}$ and field intensity $E_{x,i}^{(\text{FDE})}$ are determined at each discrete points $(w_i, h_i)$. Energy conservation is applied to connect data across sections:

$$P_{i-1} = P_i + P_{loss} \quad (M3)$$

where $P_i$ is the optical power at $i$-th cross-section, and $P_{loss}$ is the power loss during propagation. Assuming a lossless waveguide is applied to accelerate a e-beam with fixed current $I$, optical power loss is proportional to the potential difference:

$$P_{loss} = I\int_{x_{i-1}}^{x_i} E_x(x')dx'. \quad (M4)$$

Based on the Eqs. (M2)-(M3), the optical power satisfies:

$$P_{i-1} - P_i = \frac{I}{e}E_0\left\{\left[1-\left(\frac{1}{n_{\text{eff},i}}\right)^2\right]^{-\frac{1}{2}} - \left[1-\left(\frac{1}{n_{\text{eff},i-1}}\right)^2\right]^{-\frac{1}{2}}\right\}. \quad (M5)$$

Although the *real* field intensity is temporarily unknown, the optical power of each section can be calculated term by term relying on *FDE* result. Therefore, Field normalization relies on *FDE* results:

$$E_{x,i} = E_{x,i}^{(\text{FDE})}\sqrt{\frac{P_i}{P_i^{(\text{FDE})}}}. \quad (M6)$$

All the data that we need have been obtained through Eqs. (M5)-(M6), summarized in Table 1.

Since the calculations are performed at discrete points, we approximate the integral in Eq. (M2) using the trapezoidal rule, yielding the following explicit formula for spacing between cross-sections:

$$E_0\left[1-\left(\frac{1}{n_{\text{eff},i}}\right)^2\right]^{-\frac{1}{2}} - E_0\left[1-\left(\frac{1}{n_{\text{eff},i-1}}\right)^2\right]^{-\frac{1}{2}} = \frac{e}{2}(x_i - x_{i-1})(E_{x,i} + E_{x,i-1}). \tag{M7}$$

Consequently, the position $x_i$ of each section is derived by adding up the distances step-by-step.

$$x_i = x_0 + \sum_{j=1}^{i}(x_j - x_{j-1}). \tag{M8}$$

Utilizing the discrete dataset $(w_i, h_i, x_i)$, we reconstruct the waveguide's profile $h(x)$ and $w(x)$. By solving the inverse problem, we apply a simple algorithm to generate a phase-matched waveguide that allows electrons to accelerate continuously at the highest gradient within the TM-mode leakage field.

**Table 1** Basic dataset of simulation

| Data index | 0 | 1 | 2 | …… | i | …… | m |
|---|---|---|---|---|---|---|---|
| Width | $w_0$ | $w_1$ | $w_2$ | …… | $w_i$ | …… | $w_m$ |
| Thickness | $h_0$ | $h_1$ | $h_2$ | …… | $h_i$ | …… | $h_m$ |
| Effective index | $n_{\text{eff},0}$ | $n_{\text{eff},1}$ | $n_{\text{eff},2}$ | …… | $n_{\text{eff},i}$ | …… | $n_{\text{eff},m}$ |
| Field amplify (normalized) | $E_{x,0}$ | $E_{x,1}$ | $E_{x,2}$ | …… | $E_{x,i}$ | …… | $E_{x,m}$ |
| Position (TBD) | $x_0 = 0$ | $x_1$ | $x_2$ | …… | $x_i$ | …… | $x_m$ |

**Figures:**

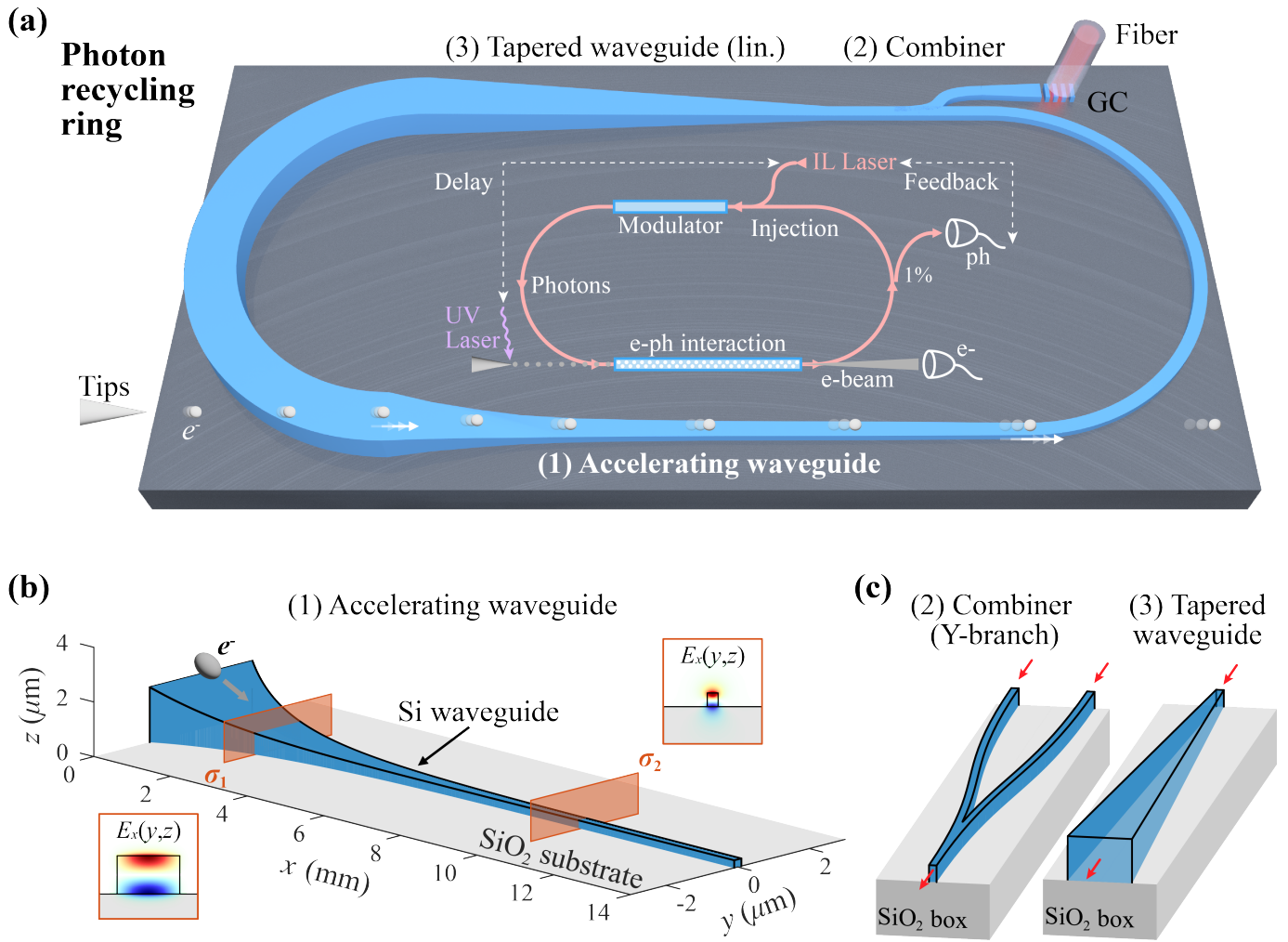

**Fig. 1 | Principle of photon-recycling racecourse with inverse-designed waveguide accelerator. (a)** Schematic diagram of the photon-recycling racecourse. A 1550 nm wavelength laser is coupled into the waveguide via a grating coupler from an optical fiber and then injected into the photon-recycling DLA system. The light undergoes mode reshaping through a linear-tapered waveguide before being guided through a curved waveguide. The reshaped light is injected into the accelerating waveguide, where the leakage optical field facilitates electron acceleration. The entire structure is fabricated on a Silicon-On-Insulator (SOI) platform. The inset illustrates the complete set-up of the system, including the laser source, electron source, modulator, and detector. **(b)** A 3D illustration of waveguide accelerator. The waveguide features nonlinear tapering in both width and/or thickness, creating a "slide" for electron beams. The longitudinal electric field component, $E_x(y,z)$, for the fundamental transverse magnetic mode (TM), is shown at two cross-sections $\sigma_1$ and $\sigma_2$. **(c)** Schematic of combiner and linear-tapered waveguide. Two Bessel-curved waveguides form a Y-shaped combiner, merging two incident waveguides into one output. The linear-tapered waveguide alters the guiding wave patterns, transitioning from a narrow profile at the combiner to a broader profile required for phase-matching and photon-recycling.

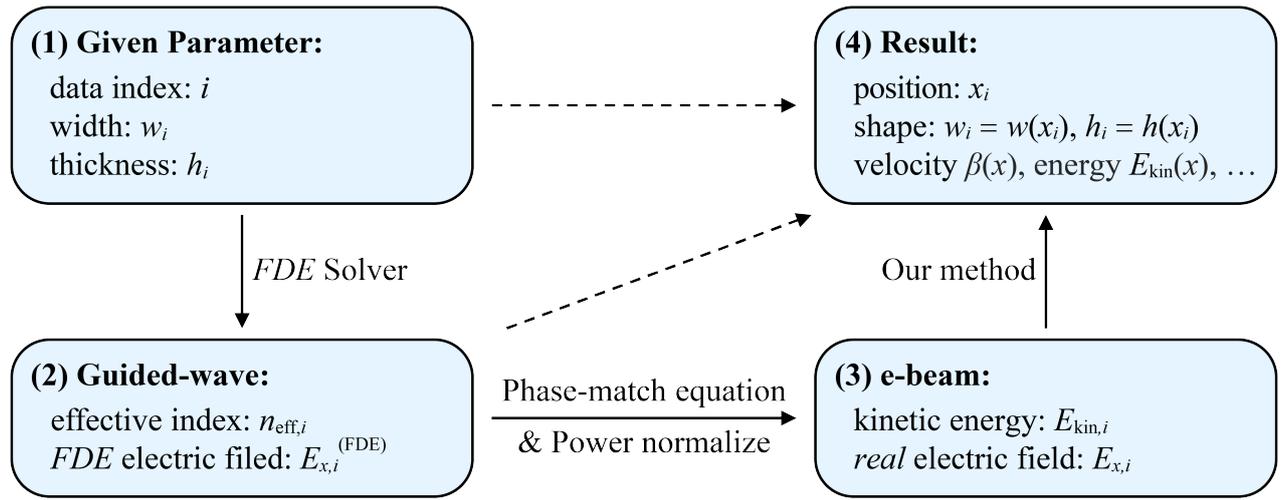

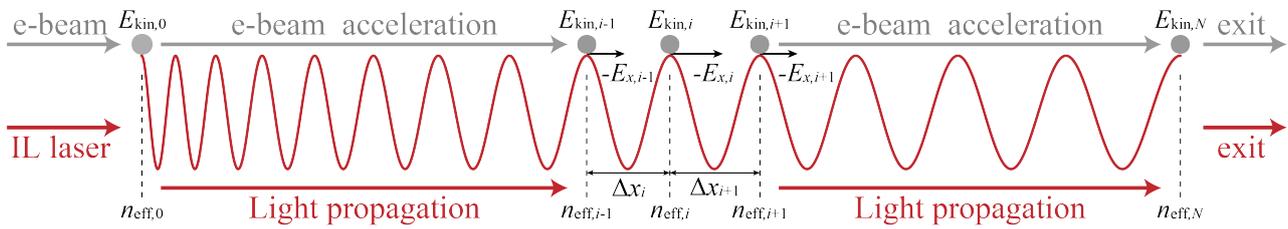

**Fig. 2 | Chart flow of the inverse-design process.** The top flowchart illustrates the entire calculation process, including the parameters and data from setting the initial parameters to obtaining the final results. The bottom schematic shows how the parameters in our optimization method are iteratively updated step by step. We begin by constructing a discrete dataset $(w_i, h_i)$ to represent the cross-sectional profiles of silicon waveguides. Using a Finite Difference Eigenmode (FDE) Solver, we solve for the $TM_0$ eigenmode for each cross-section, recording the effective refractive index, electric field intensity, and corresponding optical power. Based on the phase-matching condition, the effective refractive index is mapped to the associated electron kinetic energy. By adjusting the simulated field intensity, the optical power across different cross-sections adheres to the principle of energy conservation, enabling to compute the actual electric field intensity.

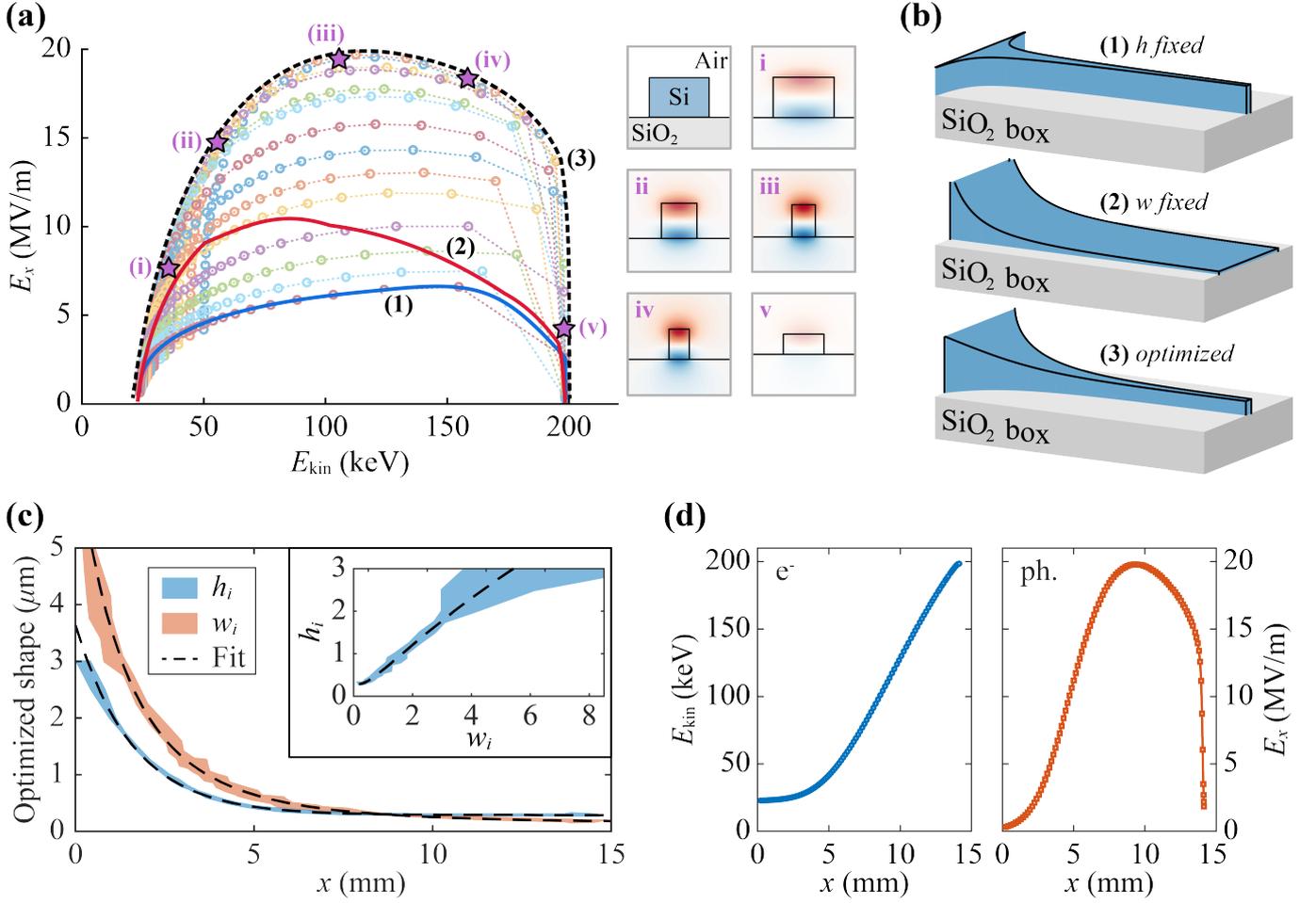

**Fig. 3 | Parameter sweep for seeking waveguide accelerator optimization. (a)** Acceleration properties $(E_{kin}, E_x)$ for waveguides of various sizes. Markers of different colors represent different waveguide sizes, and markers of the same color connected by dotted lines indicate waveguides with the same height but varying widths. Using this data, we can connect the markers to create specific types of waveguide accelerators. The blue line connects markers with the same height, representing a "trumpet" waveguide, while the red line connects markers with the same width, representing a "slide" waveguide. All the data points are enclosed by a black dashed line, corresponding to an "envelope waveguide." The basic waveguide structure used in simulations is shown in the first subgraph on the right, while typical electric field distributions ($E_x$), indicated by purple stars, are shown in the other subgraphs. **(b)** Structure diagram of trumpet waveguide, slide waveguide and envelope waveguide, from top to bottom. **(c)** Optimized shape of envelope waveguide. The envelope waveguide shrinks in both width and height, eventually transitioning from a wide, squat waveguide to a narrow, elongated one. The inset shows the relationship between width and height. The black dashed line represents the fitting result. As the width increases, the height of the envelope waveguide increases linearly. However, the fitting results suggest that the height will eventually reach a limiting value. **(d)** Changes in kinetic (left) and electric field (right) during interaction. The maxima electric field intensity is near 20 MV/m with a 100 mW photon injection.

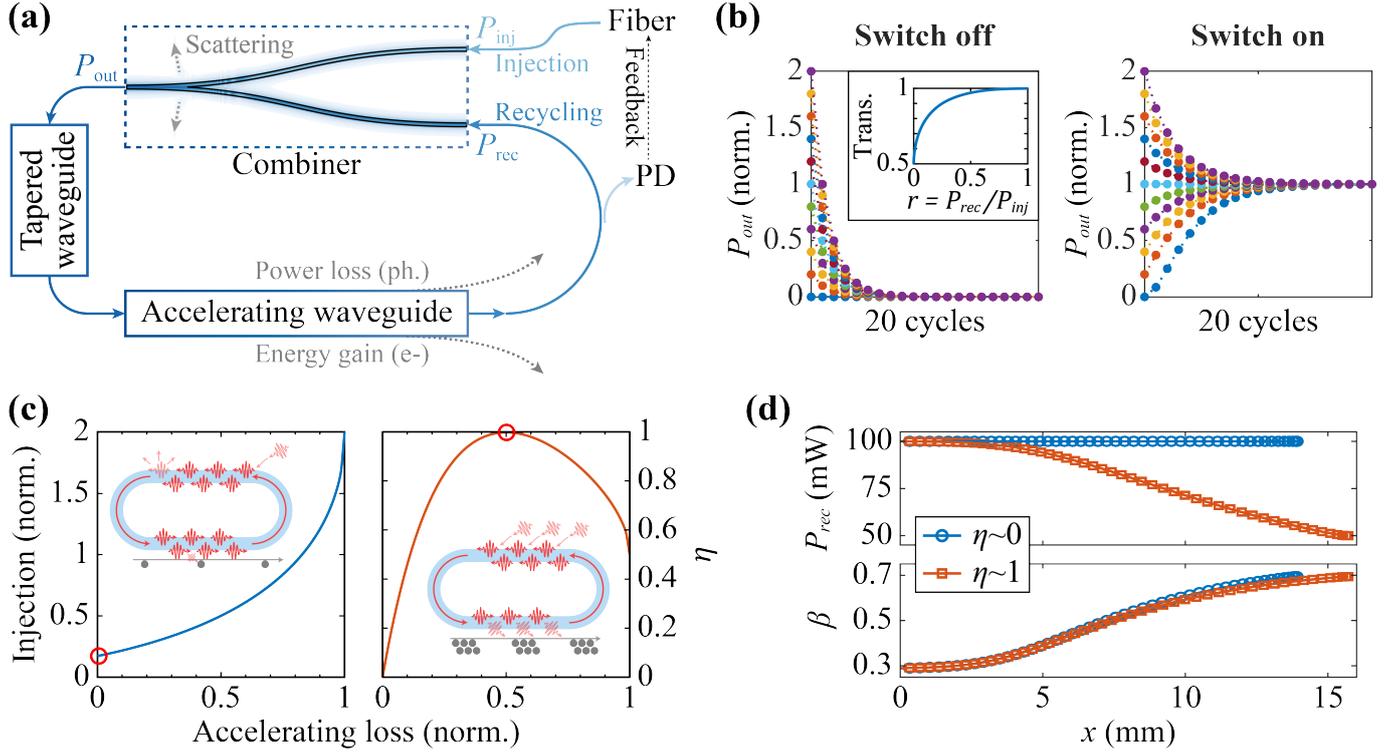

**Fig. 4 | Photon recycling in racecourse DLA system. (a)** Schematic diagram of photon intensity changes in the racecourse. Photons are continuously injected into the racecourse and combined with recycled photons at the Y-branch. The output is modulated through a linear-tapered waveguide before being injected into the accelerating waveguide. Some photons interact with electrons and are absorbed, while the remainder is recycled for the next cycle. **(b)** Transmission characteristics of combiner. The Y-branch merges injected light through nonlinear superposition (inset), which depends on the intensity ratio of the two input ports. The transmission property allows Y-branch function as a switch. When turning off the injection, the light intensity in the loop decays exponentially, terminating acceleration. When light is continuously injected with fixed power, the intensity stabilizes to a steady state within 20 cycles, regardless of the initial conditions. **(c)** Variation of photon efficiency with acceleration loss. With low acceleration losses, the loop system resonate with the continuous low-power injected light, inducing a strong electric field to accelerate individual electrons at the maximum achievable gradient. Increasing the electron beam current results in higher acceleration losses; however, an optimal level of loss enhances photon utilization. Under the assumption of lossless propagation, up to 99.8% of injected photons interact with the electron beam, thereby maximizing acceleration efficiency. **(d)** Changes in recycled power (upper) and electron relative velocity (lower) along the acceleration waveguide. Red markers represent the high-grad acceleration waveguide ($\eta \sim 0$). Blue markers represent the high-efficiency acceleration waveguide ($\eta \sim 1$).

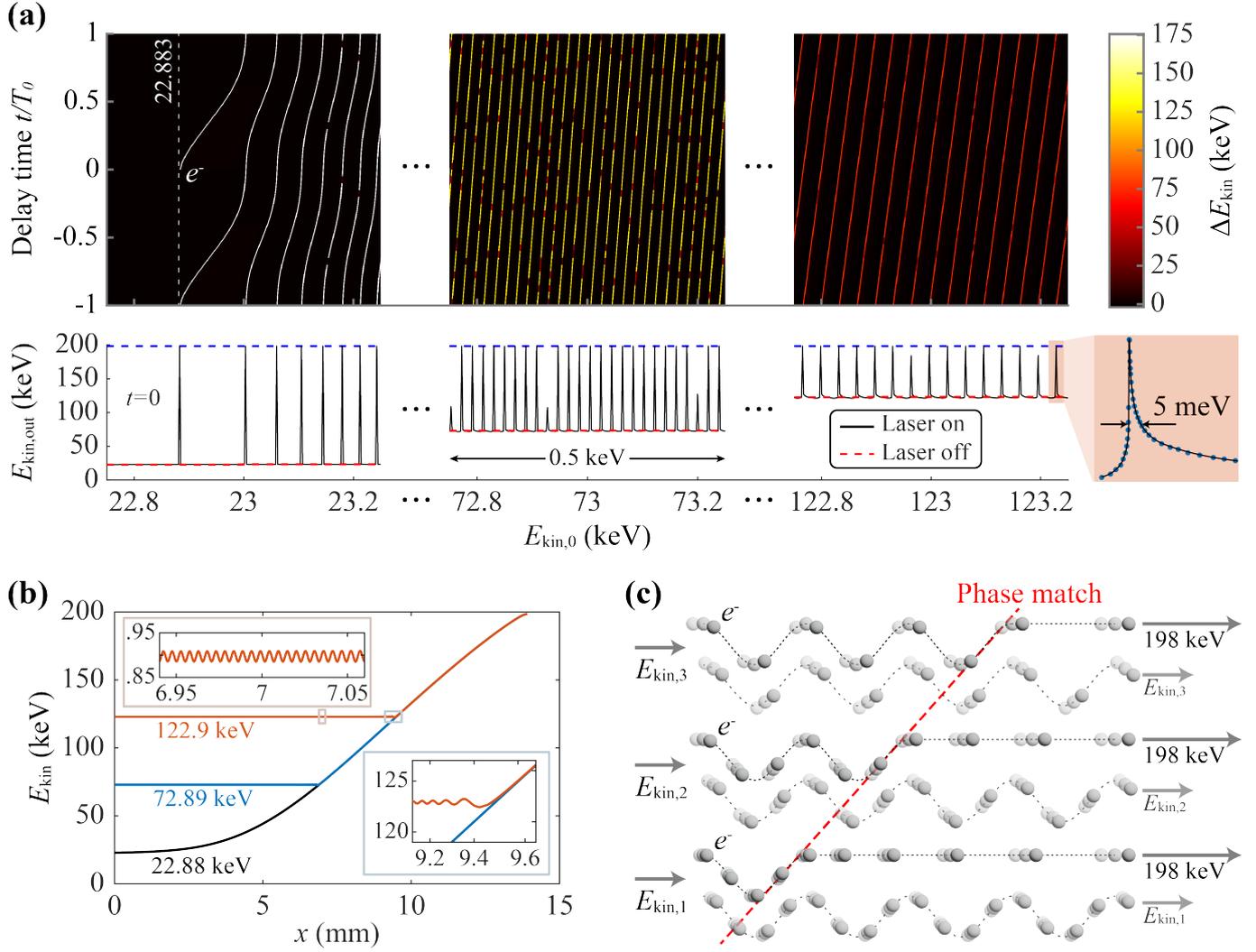

**Fig. 5 | Spectral analysis of accelerated electrons in the waveguide accelerator.** (a) Energy gains (top) and exit energies (bottom) for electrons with varying initial energies and delay times. Delay time refers to the time difference between an incoming electron and the zero-phase point of the guiding wave. The designed waveguide provides an interaction length exceeding 10 mm, producing three distinct features in the energy gain spectrum. First, the spectrum demonstrates narrow resonance peaks, each with a width of less than 10 meV. Second, the resonance peaks are distributed over a wide range (>100 keV in (a)), but they consistently occur at energies above 22.883 keV. Third, as the delay time increases, the positions of resonance peaks shift to higher energies (a blue shift). (b) The dynamic calculations for three selected peaks from (a), with an initial time delay t = 0. (c) Schematic diagram of the acceleration process for phase-matched and phase-mismatched electrons. A few electrons eventually achieve phase matching with the guiding wave after wandering to a specific position along the waveguide accelerator. Once phase-matched, these electrons continue to accelerate within the electric field, gaining energy before exiting the acceleration region. In contrast, the majority of electrons fail to achieve phase matching, resulting in no net acceleration. The waveguide accelerator acts as a "band-stop filter".

# Supplementary for

# Photon-recycling dielectric laser accelerator

Changying Li[1], Li Zhang[1], Dingguo Zheng[3], Xiaoping Liu[1], Yiming Pan[1,2]

1. School of Physical Science and Technology, ShanghaiTech University, Shanghai 200031, China
2. Center for Transformative Science, ShanghaiTech University, Shanghai 200031, China
3. Department of Electrical Engineering Physical Electronics, Tel Aviv University, Ramat Aviv 6997801, Israel

# S1. Examples of how the adaptive algorithm work

To illustrate the process of constructing the dataset and computing the waveguide profile, we consider a "trumpet-shaped" silicon waveguide with parameters $h_0 = 400$ nm and $100$ nm $\leq w(x) \leq 10$ μm. The incident light is a TM mode with wavelength of 1550 nm. and a fixed power of 100 mW. This ensure that the power transferred from photons to electrons is significantly lower than the power stored in the waveguide. Thus, we obtain

$$P_{i-1} = P_i = P_0, \qquad E_{x,i} = E_{x,i}^{(\text{FDE})} \sqrt{\frac{P_0}{P_i^{(\text{FDE})}}}.$$

We analyze two slightly different datasets, Dataset 1 and Dataset 2. First, we create two sets of width points, $w_i$ and $w_j$, and simulate waveguides profiled with sizes $(h_0, w_i)$ and $(h, w_j)$ using FDE. We calculate their effective refractive indices $n_{\text{eff},i}$ and $n_{\text{eff},j}$, and the normalized field intensity $E_{x,i}$ and $E_{x,j}$. The construction of these datasets is shown in Fig. S1a, with simulation results shown in Fig. S1b. Although we used different methods to obtain the datasets $(w, n_{\text{eff}}, E_x)_i$ and $(w, n_{\text{eff}}, E_x)_j$, the trends remain consistent. For small widths, the TM mode cannot be stably supported by the waveguide, causing mode spots to localize primarily in the silica substrate layer. As a result, the effective index initially approximates 1.44, with very low field intensity. As the waveguide width increases, the mode spots gather toward the waveguide, causing the effective index to rise rapidly from 1.44 to 2.96. Due to strong binding effects in larger waveguides, the field intensity decreases significantly as the waveguide cross-section expanded.

Using the two datasets, we apply Eqs. M5-M8 to compute the distances between each waveguide cross-section. By summing these distances, we determine the positions of each cross-section. The results, shown in Fig. S1c-d, display a waveguide with a height of $h = 400$ nm and a width that approximately follows a power-law decay function $w(x) \approx w_0(x + x_0)^{-\alpha} + w_\infty$ (where $\alpha$=0.78). This waveguide design enables electron acceleration from 32 keV to 198.5 keV over a 12 mm length while preserving phase-matching conditions throughout the acceleration process. Notably, the waveguide profile is independent of its preceding sections, rendering it "modularized". For instance, to achieve a phase-matched waveguide profiled with a width between 200 nm and 2 μm, one can directly extract the segment from 0.12 mm to 3.67 mm in Fig. S1(c) (where 200 nm $\leq w(x) \leq 2$ μm) without the need for additional calculations.

# S2. Analysis of *h*-fixed waveguide and *w*-fixed waveguide

We first extract data points where $h_i = h_0$, denoted as $(h_0, w_i, n_{\text{eff},i}, E_{x,i})$. These data points are substituted into Eqs. M5-M8 to compute the position of each cross-section, yielding a trumpet-shaped waveguide characterized by $h_0$ and $w(x_i)$. Subsequently, we calculate the kinetic energy change ($\Delta E_k$), acceleration length ($\Delta x$), and average acceleration gradient ($G$) using the following expressions:

$$\Delta E_k = E_{kin,m} - E_{kin,0}, \qquad \Delta x = x_m - x_0, \qquad G = \frac{\Delta E_k}{\Delta x}.$$

By varying $h_0$, corresponding values of $\Delta E_k$ and $G$ are determined for various trumpet waveguides, with results depicted in Figure S2a-b. Statistical analysis indicates that as $h_0$ increase to 500 nm, the energy transfer sharply increases to 170 keV. Simultaneously, the acceleration length grows nearly exponentially, resulting in the acceleration gradient $G$ sharply increasing and then exponentially decaying. Additionally, a 100 mW, 1550 nm laser input enables a trumpet waveguide with $h_0 = 300$ nm to accelerate electrons from 44.3 keV to 198.5 keV over 9.47 mm, achieving the highest acceleration gradient of 16.3 MeV/m among all designs.

Applying a similar methodology, we analyze data where $w_i = w_0$ for the slide-shaped waveguide. Interestingly, the slide waveguide exhibits acceleration characteristics nearly identical to those of the trumpet waveguide. However, the peak acceleration of the slide waveguide occurs at $w_0 = 224$ nm, with corresponding energy exchange of 160 keV and acceleration length of 10.76 mm. In contrast to the trumpet waveguide, the acceleration gradient of the slide waveguide is consistently higher. Specifically, a slide waveguide with $w_0 = 5$ μm achieve an acceleration gradient approximately 6.5 times greater than that of a trumpet waveguide with $h_0 = 5$ μm, despite both waveguides having a similar change in cross-sectional area from 5×10 μm² to 5×0.05 μm².

The similarity observed in the statistical results is closely related to the properties of guiding waves in the waveguide. When one dimension of the waveguide becomes too small, for example, $w_0 = 100$ nm, the waveguide exhibits weak confinement of the waveguide mode. This causes a significant portion of the light field to leak into the $SiO_2$ box, leading to a more dispersed mode and reduced field intensity in the air cladding. Furthermore, the leakage field within the box results in the effective refractive index of the mode being only slightly higher than that of the box material, severely limiting the acceleration range. These factors collectively make small-sized waveguides less optimal for electron acceleration. As waveguide profile increases, the guiding wave field rapidly concentrates within the waveguide, compressing the mode size and enhancing the leakage field intensity in the air cladding, which in turn results in a rapid increase in the acceleration gradient. However, when one dimension become excessively large while the other remains relatively small, such as $h_0 = 5$ μm, the mode is strongly confined within the waveguide, and the effective mode area approaches the total cross-sectional area of the waveguide. This leads to a relatively low average energy and field strength under the same power conditions, resulting in lower acceleration gradients for large-sized waveguides.

The observed differences in the statistical results are attributed to polarization effects. For TM polarization,

the leakage field primarily resides on the top and bottom surfaces of the waveguide and is more sensitive to the longitudinal structure of the waveguide. In contrast, to compressing $w$, reducing $h$ effectively "squeezes" the mode field out of the waveguide, thereby increasing the field strength of the leakage at the top surface. Consequently, "short and fat" waveguides typically exhibit higher acceleration than "tall and thin" waveguides of the same cross-sectional area. This suggested that the waveguides shown in Fig. S2d generally achieve higher acceleration than those depicted in Fig. S2b.

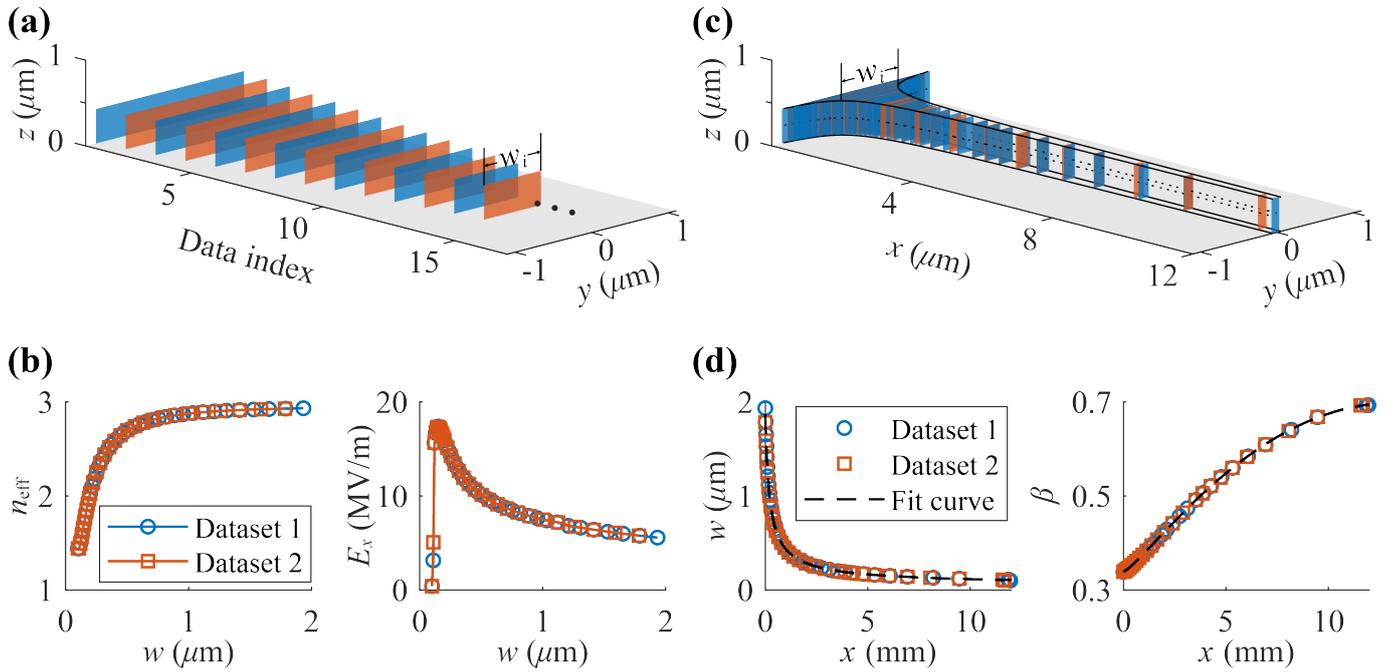

**Fig S1 | Examples of calculating methods for designing waveguide accelerator. (a)** Sketch of two different datasets $(h_0, w_i)$, where constant thickness $(h_0)$ and varying widths $(w_i)$. The same index corresponds to different widths across datasets, with different colors and markers denoting the established datasets. **(b)** Plots of the effective index (left) and electric field intensity (right) as functions of waveguide width, showing consistent trends across all datasets. **(c)** Schematic of the phase-matched waveguide designed for electron beam acceleration, illustrating the calculated the positions ($x$) of each waveguide section based on specified widths and thickness. **(d)** Plots of the waveguide width (left) and e-beams velocity $v=c\beta$ (right) as functions of position ($x$), with consistent results across all datasets.

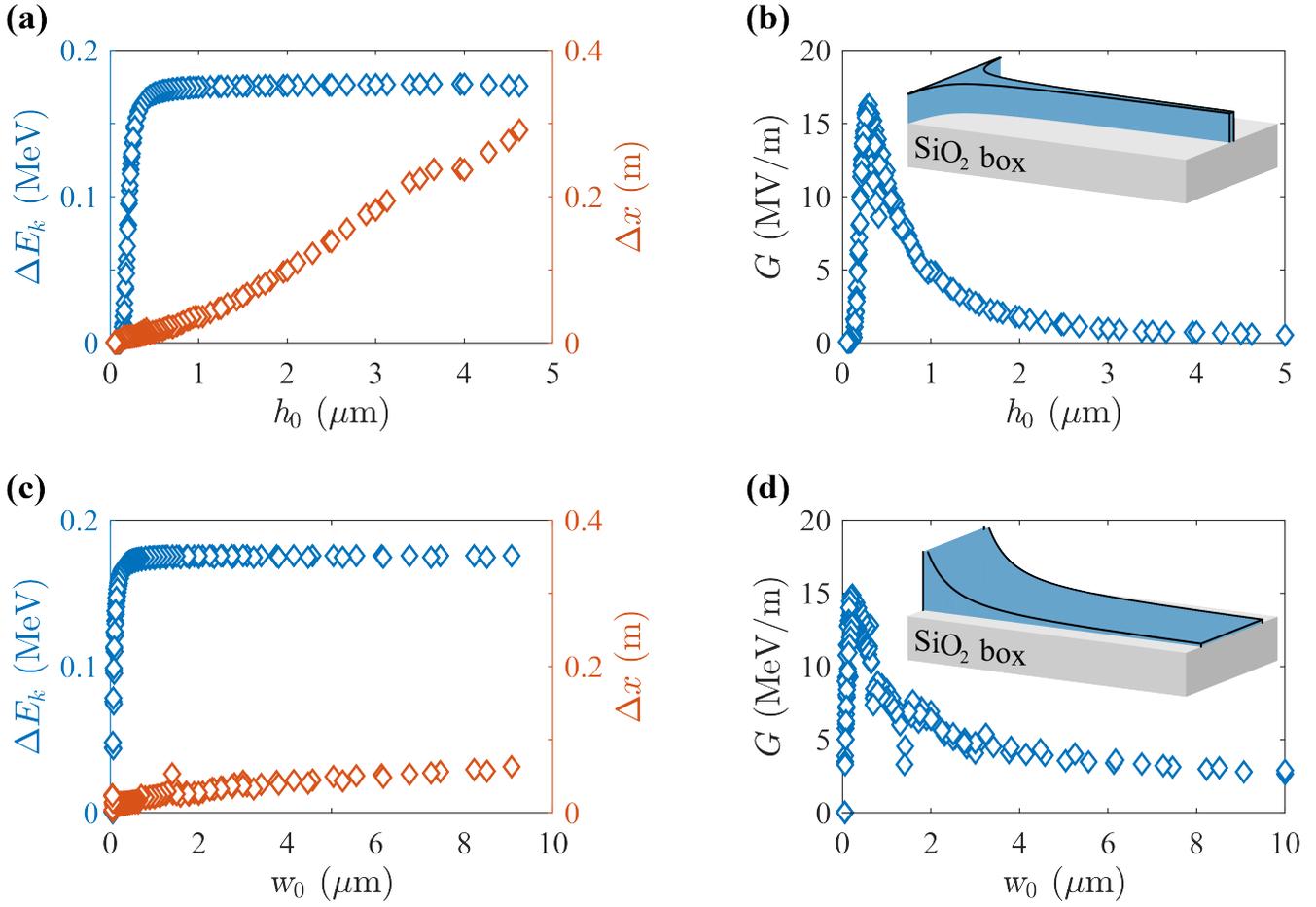

**Fig S2 | Statistics Analysis of h-fixed waveguide and w-fixed waveguide accelerator.** **(a)** Kinetic energy change (blue) and acceleration length (orange) of *h*-fixed waveguide at different thickness $h_0$. **(b)** Average acceleration gradient for *h*-fixed waveguide accelerator in terms of $h_0$, with an inset showing the *h*-fixed waveguide (aka, "trumpet waveguide"). **(c)** Kinetic energy change (blue) and acceleration length (orange) for *w*-fixed waveguide in terms of $w_0$. **(d)** Average acceleration gradient for *w*-fixed waveguide in terms of $w_0$, with an inset depicting the *w*-fixed waveguide (aka, "slide waveguide"). Both waveguide profiles exhibit similar acceleration characteristics, with the slide waveguide demonstrating a higher average acceleration gradient at large size.